# Aging, memory, and nonhierarchical energy landscape of spin jam

Anjana Samarakoon[1], Taku J. Sato[2], Tianran Chen[1], Gai-Wei Chern[1], Junjie Yang[1], Israel Klich[1], Ryan Sinclair[3], Haidong Zhou[3], and Seung-Hun Lee[1*]

[1]Department of Physics, University of Virginia, Charlottesville, VA 22904, USA

[2]Institute of Multidisciplinary Research for Advanced Materials, Tohoku University, Katahira, Sendai 980-8577, Japan

[3]Department of Physics and Astronomy, University of Tennessee, Knoxville, TN 37996, USA

*Corresponding author (email: shlee@virginia.edu)

**Abstract:** The notion of complex energy landscape underpins the intriguing dynamical behaviors in many complex systems ranging from polymers, to brain activity, to social networks and glass transitions. The spin glass state found in dilute magnetic alloys has been an exceptionally convenient laboratory frame for studying complex dynamics resulting from a hierarchical energy landscape with rugged funnels. Here, we show, by a bulk susceptibility and Monte Carlo simulation study, that densely populated frustrated magnets in a spin jam state exhibit much weaker memory effects than spin glasses, and the characteristic properties can be reproduced by a nonhierarchical landscape with a wide and nearly flat but rough bottom. Our results illustrate that the memory effects can be used to probe different slow dynamics of glassy materials, hence opening a window to explore their distinct energy landscapes.

**Main Text:** If the energy landscape of a system resembles a smooth vase with a pointy bottom end, upon cooling the system goes quickly into the lowest energy state, i.e., the global ground state that is usually associated with crystalline order. If the energy landscape is more complex with many metastable states, i.e., local minima, then cooling may lead the system into local minima resulting in a glassy order. The concept of such energy landscapes has been instrumental in explaining the glassiness that is ubiquitous in a wide range of systems, including atomic clusters (1), structural glasses (2, 3), polymers (4), brain activity (5), and social networks (6). Several different topological types of energy landscapes were proposed to characterize different glassiness and the associated slow dynamics (7, 8). For instance, a rugged funnel-shaped landscape shown in

Fig. 1A was proposed to understand the physics of biopolymers (9, 10) and dilute magnetic alloys called spin glass (11).

Magnetic glass systems (12–16) present a unique opportunity to microscopically study the relation between the energy landscape and low temperature properties. The most studied magnetic glass state is the conventional spin glass realized in dilute magnetic alloys such as $Cu$Mn and $Au$Fe. Here, dilute magnetic ions (Mn and Fe) in a nonmagnetic metal interact via the long-range Ruderman–Kittel–Kasuya–Yosida (RKKY) interaction whose magnitude and sign change with distance between the randomly placed magnetic ions (11). The randomness drives the system into the spin glass state below a critical temperature, $T_f$, that is comparable to the mean-field magnetic energy scale, i.e., the absolute value of the Curie–Weiss temperature, $|\Theta_{CW}|$. For instance, for $Cu$ − 2 at. % Mn ($Cu$Mn2% hereafter), $\Theta_{CW} = -45$ K and $T_f = 15.5$ K. Another distinct glassy state called a spin jam has been recently suggested to appear in densely populated frustrated magnets (17–21). At the mean-field level, these systems are expected to remain in a classical spin liquid down to absolute zero temperature, due to macroscopic classical ground state degeneracy. Quantum fluctuations, however, lift the degeneracy and lead the system to the spin jam state below $T_f$ that is much lower than $|\Theta_{CW}|$. For instance, for $SrCr_{9p}Ga_{12-9p}O_{19}$ (SCGO)($p = 0.97$), $\Theta_{CW} = -500$ K and $T_f = 3.8$ K. The stark different ratios of $|\Theta_{CW}|/T_f$ for the spin jam and spin glass suggest that the two states might have qualitatively distinct energy landscapes.

Aging and memory effects have been key features of glassy systems due to the intrinsic slow dynamics. The thermo-remanent magnetization (TRM) method is the most effective way so far to investigate these effects (12–14); for the measurements, the sample is first cooled down from well above $T_f$ to base temperature with a single stop for a waiting time, $t_w$, at an intermediate temperature $T_w$, under zero field. While waiting at $T_w$, if the system has as many nearly degenerate metastable states at low energies as spin jam and conventional spin glasses have, the system will relax to the accessible lower energy states than when no waiting is imposed. The longer $t_w$ is, the lower energy states the system will relax to, which is called "aging". Once cooled down to base temperature, the TRM is measured by applying a small field of a few gauss upon heating at a constant rate. During the measurements, when the temperature approaches the temperature of aging, $T_w$, the system revisits the lower energy states reached during the wait time that are associated with the energy scale of $k_B T_f$, where $k_B$ is the Boltzmann constant. Upon further heating, the system goes to higher energy states allowed within $k_B T$. This is referred to as the aging and memory effect.

We have performed the TRM measurements on two spin jam prototypes, SCGO($p = 0.97$) and $BaCr_{9p}Ga_{12-9p}O_{19}$ (BCGO)($p = 0.96$) in which the magnetic $Cr^{3+}$ ($3d^3$) ions form a highly frustrating quasi-2D triangular network of bipyramids (17–21) and a spin glass prototype $Cu$Mn2% in which the 2% low concentration of the magnetic Mn atoms is embedded in the nonmagnetic Cu metal. Strong aging and memory effects have been observed in $Cu$Mn2%, whereas the effects are much weaker in SCGO and BCGO. Fig. 2 A–C shows the TRM data obtained from SCGO($p = 0.97$), BCGO($p = 0.96$), and $Cu$Mn2%, respectively, with several different values of $t_w$ ranging from 6 min to 100 h, at $T_w = T_f \sim 0.7$. All samples exhibit similar aging and memory effects that increase with increasing $t_w$. These indicate the existence of numerous metastable states and slow dynamics in all systems. Despite the similarity, there is a clear difference: For the $Cu$Mn2% magnetic alloy, considerable aging occurs at $T_w$ even for a short $t_w$ of 6 min (data in violet in Fig. 2C), whereas for the spin jam SCGO($p = 0.97$) and BCGO($p = 0.96$), there is very small aging for $t_w = 6$ min (data in violet in Fig. 2 A and B). Furthermore, in the case of $Cu$Mn2%, as $t_w$ increases, the memory effect increases to develop a dip at $T_w$ for $t_w \geq 3\ h$. On the other hand, for SCGO($p = 0.97$) such a dip never appears even for $t_w = 100$ h; instead only a weak memory shoulder appears.

The memory effect can be quantified by the aging-induced relative change in the magnetization, $(M_{aging} - M_{ref})/M_{ref}$, where $M_{aging}$ and $M_{ref}$ are the magnetization with and without aging, respectively. Fig. 2D shows $(M_{aging} - M_{ref})/M_{ref}$, measured at $T_w/T_f \sim 0.7$ for SCGO($p = 0.97$) (solid symbols), BCGO($p = 0.96$) (symbols with a line), and CuMn2% (open symbols), as a function of $t_w$. These data are consistent with a previous study on SCGO($p = 0.956$) with $t_w$ up to 5.83 h (13). In the case of CuMn2%, as $t_w$ increases from 6 min to 100 h, $(M_{aging} - M_{ref})/M_{ref}$ continues to gradually increase from 3.4% to 8.2%. On the other hand, for SCGO($p = 0.97$), $(M_{aging} - M_{ref})/M_{ref}$ increases gradually from 0.6% to 2.4%, and for BCGO($p = 0.96$), from 0.7% to 3.1%, as $t_w$ increases from 6 min to 10 h. The increase rate of $(M_{aging} - M_{ref})/M_{ref}$ seems to decrease for $t_w > 10$ h, reaching 2.7% at $t_w = 100$ h for SCGO($p = 0.97$). We emphasize that over this wide range of $t_w$ up to 100 h the susceptibility curve is always monotonically dependent on temperature up to the freezing point (Fig. 2 A and B), in sharp contrast to CuMn2%.

Fig. 3 shows the memory effect for various values of $0.4 \lesssim T_w/T_f \lesssim 1$ measured with $t_w = 10$ h. All systems exhibit maximal memory effect for $T_w/T_f \sim 0.7$. When $T_w/T_f$ increases or decreases from the maximal value, then the memory effect becomes weaker. The weakening, however, is more rapid in CuMn2% than in SCGO and BCGO; for CuMn2%, $(M_{aging} - M_{ref})/M_{ref}$ decreases from 7.2% for $T_w/T_f \sim 0.7$ to 2.7% for $T_w/T_f \sim 0.9$, whereas for SCGO($p = 0.97$) [BCGO($p = 0.96$)], $(M_{aging} - M_{ref})/M_{ref}$ decreases from 2.4% (3.1%) for $T_w/T_f \sim 0.7$ to 1.7% (2.5%) for $T_w/T_f \sim 0.9$. The pronounced memory effects found in CuMn2% may hint at an energy landscape with a more hierarchical structure. On the other hand, the weak memory effect, observed in SCGO and BCGO, which is uniform over a wide range of $0.4 \lesssim T_w/T_f \lesssim 1$, suggests an energy landscape with a less hierarchical structure.

Rejuvenation and memory effects have proved difficult to reproduce in standard simulations of supercooled liquids or spin glasses, due to the large phase space to be covered and large spread of time scales involved (22, 23). Several successful attempts were made, such as a multilayer random energy model (24) and a model of thermally activated number sorting (23).

None of the studies, however, investigated how different topologies of the energy landscape will impact the memory effects. Here we have done so by taking a phenomenological approach based on a multilayer energy model. As shown later, this approach reproduces qualitatively the differences between memory effects associated with different landscapes.

We performed Monte Carlo simulations on two types of energy landscapes suggested for the spin glass and spin jam. Although the energy surface in both cases is characterized by numerous local minima, the distribution and connectivity of these minima are very different. Here we adopt the so-called barrier tree representation (8, 25, 26) in which the local minima correspond to leaves of the tree, whereas the branching points denote the barriers separating disconnected valleys and/or minima. Details can be found in Fig. S1 and discussion in Supporting Information.

Fig. 1A shows a funnel-type barrier tree that is characteristic of the conventional spin glass. A rugged funnel here corresponds to a single long branch (the global minimum) with many dead branches splitting from it (8, 25). The experimentally observed memory effect is intimately related to a multitude of energy and time scales in the low-energy configurational space. For the funnel-type landscape, a hierarchical structure of energy scales is encoded in the different levels of the barrier tree. The energy barriers $\varepsilon_l$ at level $l$ are characterized by a temperature $T_l$ such that $T_1 > T_2 > \cdots > T_L$, where $L$ is the number of levels of the tree (24). The freezing temperature is $T_f \approx T_1$. The relaxation of the system in this hierarchical structure exhibits complex temperature-dependent dynamics. Typically, because the relaxation time at level $l$ scales as $\tau_l \sim \tau_0 e^{\varepsilon_l/T}$, where $\tau_0$ is a microscopic time scale, the relaxation dynamics start to show exponential slowing down at level $l$ when $T < T_l$. Depending on the population of dead-end local minima at each level, the system fluctuates over a small window of levels determined by $T_w$ in the experiments. A longer tw at this temperature allows the system to relax to a deeper and larger (entropically) valley of the energy surface. The memory effect observed during the reheating process results from the fact that the system is trapped in this special landscape basin. The susceptibility, $\chi_{DC}$, as a function of temperature is shown in Fig. 4 A–C for three different $T_w$. The DC susceptibility computed using a random magnetization model (24) shows a clear dip that depends on $T_w$ as well as $t_w$. In particular, a longer $t_w$ gives rise to a larger susceptibility reduction. It should be noted that other contributions to $\chi_{DC}$ such as continuous spin fluctuations are not included in the landscape tree dynamics simulations.

In contrast to ordinary spin glass, the energy structure in a spin jam results from quantum and classical fluctuations breaking an exactly flat landscape (20). Importantly, the energy scale for glass transition $T_f$ is determined by the fluctuations and is two orders of magnitude smaller than the Curie–Weiss temperature (20). We expect the resulting landscape to feature broad basins and within each basin numerous microstates, as shown in Fig. 1B. As the local minima in spin jam result from the original zero energy mode of the classical spin Hamiltonian, it is plausible that the energy minima here are clustered into different branches (labeled by m) each characterized by a different energy scale $T_m$. For a particular cluster or branch of minima, the temperature $T_m$ underscores the energy barrier due to quantum fluctuations. $T_m$ is a random variable and is uniformly distributed in the interval of $[0, T_f]$. $T_w$ sets a threshold such that clusters with $T_m > T_w$ exhibit slow relaxation dynamics, whereas a longer $t_w$ helps the system find the cluster with a lower overall energy and larger entropy. This property underscores the weak memory effect observed in spin jam. Again, the fact that the system is trapped in this special cluster manifests itself as the memory effect during rewarming. The simulated susceptibility of the nonhierarchical tree, shown in Fig. 4 D–F, shows a memory effect that depends on both $T_w$ and $t_w$, similar to the spin glass. However, the salient feature is rather different: Contrary to the narrow dip in the hierarchical tree that appears even for short waiting time $t_w > 1$ (Fig. 4 A–C), the susceptibility here exhibits a wide shoulder-like feature over a much wider range of tw for each Tw. As shown in more detail in Fig. S2A, for $T_w$ = 0.6 $T_f$ the nonhierarchical landscape fails to yield a narrow dip over six orders of magnitude of the Monte Carlo (MC) steps. Remarkably, this finding is consistent with the experimental data revealing a shoulder-like feature for spin jam (Fig. 2 A and B) vs. the substantial dip for the ordinary spin glass (Fig. 2C). Furthermore, the functional dependence of the memory effect on waiting time is nicely reproduced for both systems in Fig. 2D over three orders of magnitude.

The picture that emerges from the bulk susceptibility and Monte Carlo simulations is that the energy landscape of a spin jam is qualitatively different from the rugged funnel-type landscape of a spin glass. The hierarchical structure allows a natural realization of multiple energy scales (e.g., ref. 27) that is crucial to the memory effect. On the other hand, the aging dynamics in the spin jam are well described by an essentially nonhierarchical barrier tree with more uniform branching. This result is consistent with the fact that the rough energy landscape in spin jam results from quantum fluctuations that lift the otherwise degenerate classical ground states. In particular, the weak memory effect at short times found in a spin jam may be interpreted as a result of the large time it takes the system to wander among the numerous roughly equivalent minima at a given energy scale.

The transition from a spin liquid to a spin jam in densely populated frustrated magnets upon cooling may be viewed as an effective reduction of degrees of freedom. In SCGO (20, 21) and kagome antiferromagnet (28) the origin of the reduction can be induced by quantum fluctuations. We remark that the transition bears some analogy to the transition from a structural (mechanical) liquid state to a mechanical jam by increasing the concentration of the atoms, i.e., pressure (29). Both frustrated magnets and the mechanical jam have a large number of metastable states, other than their ground states, in the vicinity of their liquid states. The configurational entropy of these states ranges from extensive, as in mechanical jams and coplanar states of the kagome antiferromagnet (28, 30), to subextensive as in the locally collinear states of an ideal SCGO (20). Both types of systems are expected to feature a relatively shallow energy landscape of accessible states due to their proximity to a uniform liquid state. It is interesting to note a possibly related observation that the ensemble of metastable states in self-generated Coulomb glasses is shallow compared with more ordinary electron glasses relying on quench disorder (31).

The two fundamentally different trees studied here can be cast in the framework of complex networks (32, 33). The spin glass's hierarchical energy landscape (even with a fractal structure) resembles the so-called scale-free network (33), proposed to explain internet connections and ecological and neural networks (34). In this network, there are highly connected dominating nodes, each of which corresponds to the global minimum of a rugged funnel. On the other hand, the spin jam's nonhierarchical landscape corresponds to a network consisting of weakly connected clusters that are homogenous on a larger scale.

**ACKNOWLEDGMENTS:** Work at the University of Virginia by S.H.L. and I.K. was supported in part by US National Science Foundation (NSF) Grants DMR- 1404994 and DMR-1508245, respectively. A.S. is supported by Oak Ridge National Laboratory. Work at Tohoku University was partly supported by Grants-in-Aid for Scientific Research (24224009, 23244068, and 15H05883) from Ministry of Education, Culture, Sports, Science and Technology of Japan. Work at the University of Tennessee was supported by US NSF Grant DMR-1350002.

**Figure and Captions**

Fig. 1. (A) (Upper) Schematic energy landscape of a conventional spin glass that consists of many hierarchical rugged funnels and (Lower) the corresponding hierarchical tree representation. (B) (Upper) Schematic diagram of nonhierarchical energy landscape of the spin jam that has wide nearly flat rough bottom and (Lower) the corresponding nonhierarchical barrier tree representation.

Fig. 2. (A–C) Bulk susceptibility, $\chi_{DC} = M/H$, where $M$ and $H$ are magnetization and applied magnetic field, respectively, obtained from (A) SCGO(p = 0.97), (B) BCGO(p = 0.96), and (C) a spin glass CuMn2%, with $H$ = 3 G. Symbols and lines with different colors indicate the data taken with different waiting times, $t_w$, ranging from 0 h to 100 h, at $T_w/T_f \sim 0.7$, where $T_w$ and $T_f$ are the waiting and the freezing temperature, respectively. (D) From the data shown in A–C, the aging effect was quantified for the three systems by $(M_{ref} - M)/M_{ref}$, where $M_{ref}$ is the magnetization without waiting, and it was plotted as a function of $t_w$ in a log scale. The "+" symbols mark the results of our MC simulations. Details of the simulations can be found in Supporting Information.

Fig. 3. (A–C) $\chi_{DC}$ and (D–F) $(M_{ref} - M)/M_{ref}$ measured for (A and D) SCGO($p$ = 0.97), (B and E) BCGO($p$ = 0.96), and (C and F) CuMn2%, with $t_w$ = 10 h, at various waiting temperatures.

Fig. 4. A–C and D–F show the simulated DC susceptibility during the reheating process for the hierarchical and nonhierarchical trees, respectively, at three different waiting temperatures $T_w = 0.2T_f$ (A and D), $0.4T_f$ (B and E), and $0.6T_f$ (C and F). Different curves in each panel correspond to varying tw measured in units of the total cooling time.

Fig. 1

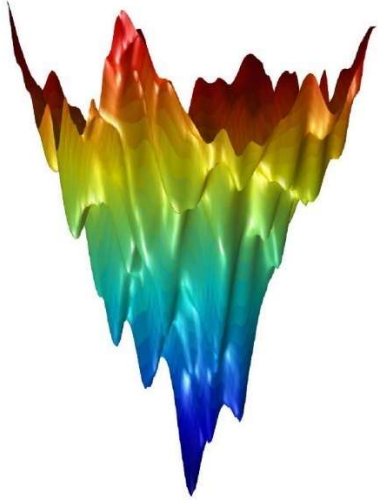

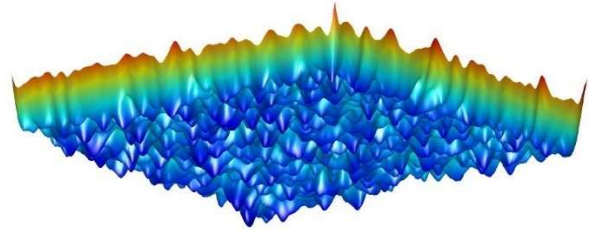

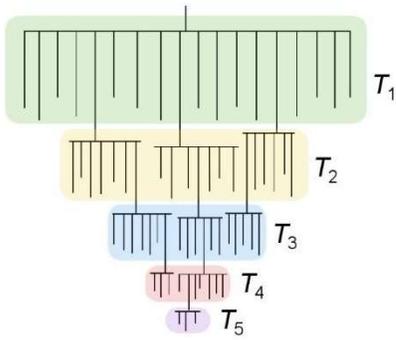

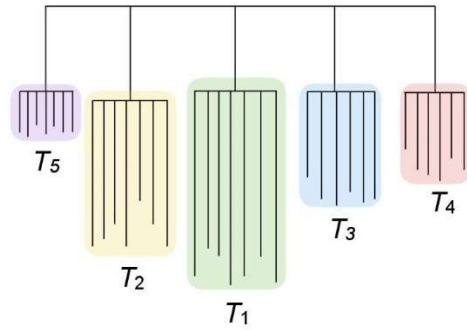

Fig. 2

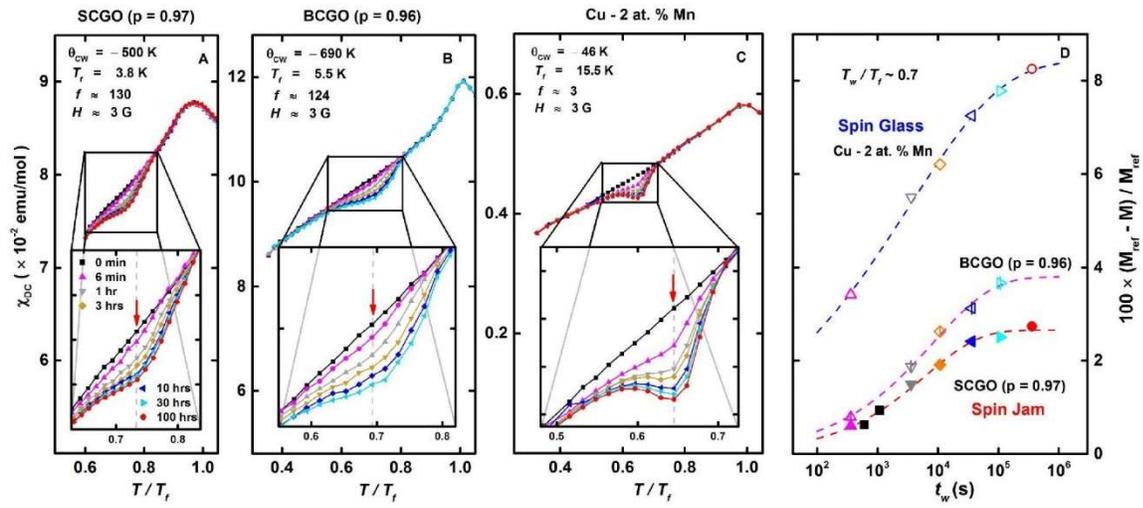

Fig. 3

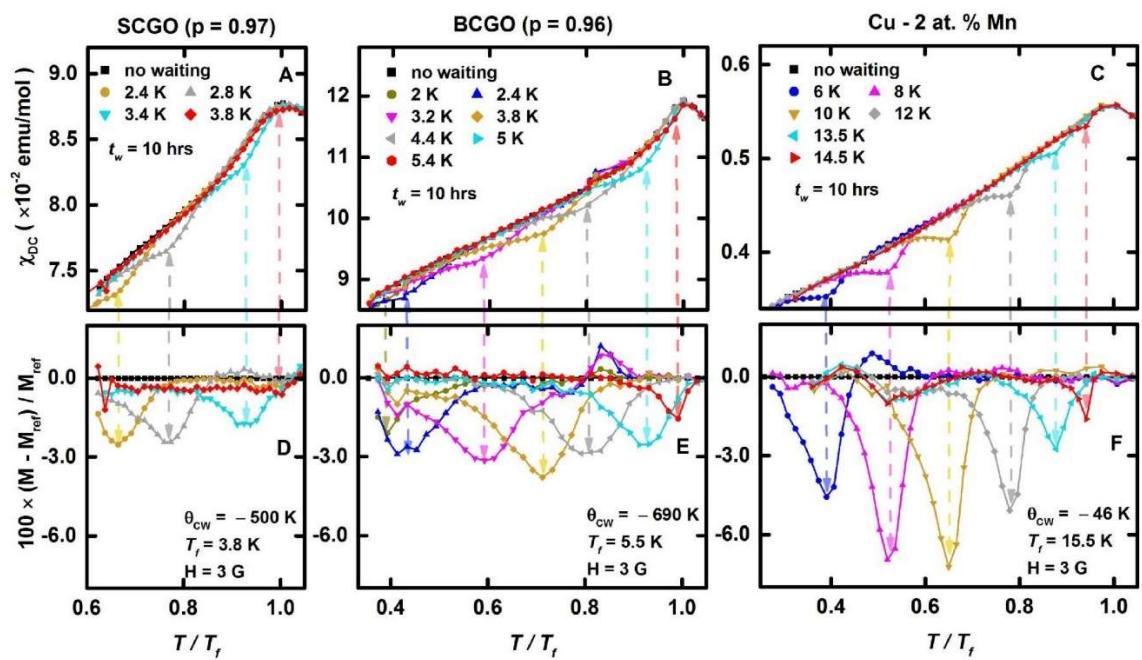

Fig. 4

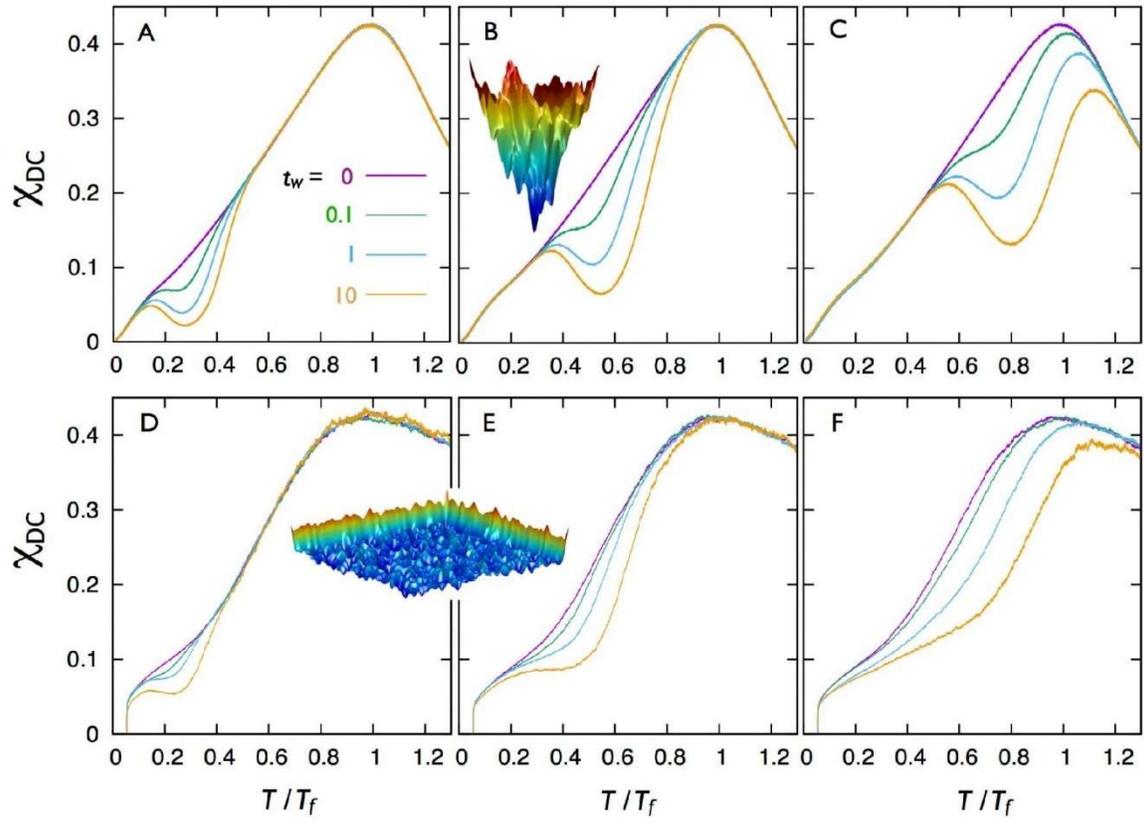

# Supporting Information

## Model and Monte Carlo Simulations

Understanding the dynamics of a glassy system via the energy landscape approach has led to promising insights into puzzling phenomena, such as temperature-dependent aging and memory effect. In this framework, the energy landscape is seen as a set of basins of attraction, and the system evolves through a succession of jumps between their local minima (35, 36). This approach focuses on the interbasin transitions without treating explicitly the fast (high-frequency) intrabasin dynamics. Over the past decades, much effort has been devoted to characterizing the structure and topology of the energy landscape for various glass-forming systems (8, 25, 37). In particular, the so-called disconnectivity graph (25) has become a widely used approach for visualizing and representing the multidimensional potential energy surface. The disconnectivity graph summarizes the local minima and saddle points of an energy landscape into a tree. Each leaf in this tree representation corresponds to a local minimum, whereas the branching point is a transition state (saddle point) connecting different local minima. Another approach to describe the energy landscape is to use the language of complex networks (33, 36).

The disconnectivity graph (also referred to as a barrier tree representation) can be constructed numerically from the database of local minima and the kinetic pathways for small molecules or lattices (37). Monte Carlo sampling is often required to construct the representative barrier tree for a larger system. These approaches require microscopic details of the system at hand, which are hard to process for complicated physical systems. An alternative, phenomenological, approach for complex systems uses a statistically based characterization of the barrier tree, which is the method adopted in our work. For a given energy landscape representation, either through a barrier tree or through a complex network, the dynamics of the system can be simulated as a random walk on the tree or the network. A master equation is often used to study the resultant dynamics (25, 35, 36). Here we use the Markov chain Monte Carlo simulations coupled with a dynamical tree method to study the memory effect of spin glass and spin jam states. Our Monte Carlo approach offers the advantage of being applicable to barrier trees with complex structures without the need of introducing further approximations as in the master equation method.

In our simulations, the relaxation dynamics of the system are modeled as a random walk on the barrier tree. Each node of the tree, corresponding to either a local minima or a saddle point, represents a specific microscopic spin configuration. Transition between two nodes corresponds to modifying a small number of spins. In our simulations, the structure and properties of the barrier trees are characterized statistically. Specifically, the tree is described by a set of random numbers satisfying certain probability distributions.

We first discuss the statistical description of the barrier tree for a conventional spin glass. Here, the tree has a hierarchical structure with many levels. A node at a lower level (larger $l$) corresponds to a lower energy state (Fig. 1A). The barrier energy $\varepsilon$ at level $l$ is an independent random variable with an exponential distribution $p(\varepsilon) = e^{-\varepsilon/T_l}/T_l$ characterized by the temperature $T_l$. This construction is similar to the so-called random energy or random trap models (38, 39) that are shown to exhibit the characteristic aging behavior in spin glasses.

The characteristic temperature $T_l$ decreases with increasing levels; i.e., $T_1 > T_2 > \cdots > T_{l_{max}}$, corresponding to smaller energy barriers at the bottom of the hierarchical tree. In our simulations, we assume that the characteristic temperature $T_l \sim T_1 e^{-\alpha}$ decreases exponentially with the level index. The relaxation of the system in this hierarchical structure exhibits complex temperature-dependent dynamics. Typically, because the relaxation time at level $l$ scales as $\tau_l \sim \tau_0 e^{\varepsilon_l/T}$, where $\tau_0$ is a microscopic time scale, the relaxation dynamics start to show exponential slowing down at level $l$ when $T < T_l$. Interestingly, the progressive slowing down of the relaxation dynamics can be viewed as the fact that the system undergoes a series of glass transitions with decreasing temperature. The largest energy scale $T_1$ then determines the nominal freezing transition temperature $T_f$. This hierarchical construction is consistent with the picture of temperature-dependent energy barriers (40), which is shown to be crucial for the occurrence of memory effect in conventional spin glass.

At each level, whether the node is a local minimum (and thus a dead end) or a saddle point is specified by a constant $0 < \lambda_l < 1$; i.e., $\lambda_l$ is the probability that a given node at level $l$ is a local minimum. Another random number $n_b$ is used to specify the branching or the number of descendants of a saddle point. Finally, for the calculation of magnetic susceptibility, we use the simple random magnetization model discussed in ref. 24 for the barrier tree. Specifically, the magnetization of a particular state at level $L$ is given by $M = m_0 + m_1 + \cdots + m_L$. The magnetization contribution $m_l$ from level $l$ is a random number uniformly distributed in the interval [$-\mathcal{M}_l, \mathcal{M}_l$], where the bound $\mathcal{M}_l$ is assumed to decrease exponentially with increasing levels. Consequently, a Zeeman coupling $\mathcal{H}_z = -H \cdot M$ is included in the Monte Carlo simulations of the reheating process.

We use the standard Metropolis dynamics in our Monte Carlo simulations. Because the barrier tree is specified only statistically, there is no need to create a tree at the beginning of the dynamical simulations. Instead, we generate the barrier tree dynamically according to the desired statistical properties as discussed above. However, additional bookkeeping is required to describe a system currently at level $L$. Specifically, we need to keep track only of all of the barrier energies and magnetizations from levels $l \leq L$, i.e., $\{\varepsilon_1, \varepsilon_2, \cdots, \varepsilon_l, \cdots, \varepsilon_L\}$ and $\{m_1, m_2, \cdots, m_l, \cdots, m_L\}$. A Monte Carlo step then consists of the following procedures: (*i*) Determine whether this node is a local minimum (a dead end) or a saddle point. This can be done by generating a random number r uniformly distributed between 0 and 1. If $r < \lambda_L$, then the current state is a local minimum, and the system can move only upward to escape this local trap. (*ii*) If the current node is a saddle point, then we generate another uniformly distributed random number $r' \in [0,1]$. If $r' > 1/(n_b + 1)$, then we attempt to move the system upward. Otherwise we move the system downward to a lower level (closer to the global minimum). (*iii*) For a downhill update, we first increase the level by one.

Next, we generate new random numbers $\varepsilon_{L+1}$ and $m_{L+1}$ according to the respective probability distribution and add them to the lists of barrier energies and magnetizations, respectively. (*iv*) Finally, for an uphill update we first compute the energy cost $\Delta\varepsilon = \varepsilon_L + Hm_L$. Then a standard Metropolis criterion is used to determine whether this upward movement is accepted or not. If the uphill move is accepted, we then erase $\varepsilon_L$ and $m_L$ from the respective lists. The above procedures are illustrated in Fig. S1.

We note that the dynamical tree simulation is valid as long as the number of branchings $n_b \gg 1$ (in the simulations we took $n_b \sim 500$). Under this condition, we can neglect the possibility that the system will visit exactly the same lower-energy states more than once in our finite-time simulation. We also find that a rather large $\lambda_l$ is required to observe a noticeable memory effect. This condition simply means that there are many dead-end local traps along the way toward the global ground state, which is consistent with the rugged funnel-type energy landscape. In our simulations, we assumed a maximum number of level $l_{max} = 50$ and took $\lambda_l$ to increase linearly from 0.9 to 1 at $l_{max}$. We also note that barrier trees characterized by these statistical properties are similar to the "palm tree" pattern in the classification of disconnectivity graphs (8, 37).

The spin jam glassy state, on the other hand, is characterized by a very different energy landscape. This is because the numerous minima in the spin jam originate from quantum fluctuations that lift the otherwise flat energy surface at the classical level; we expect a uniform, nonhierarchical barrier tree structure, shown in Fig. 1B in the main text, for spin jam. The lack of hierarchical structure in this type of tree indicates that the weaker memory effect of spin jam results from a different mechanism. As the local minima in a spin jam result from the original zero energy mode of the classical spin Hamiltonian, it is plausible that the energy minima in the spin jam are grouped into clusters with different average barrier heights. This nonhierarchical tree resembles the so-called "banyan tree" pattern (8, 36). In this tree structure, different clusters are separated by a large barrier energy $T_b$, whereas the barrier energies within a cluster (labeled by $m$) are random numbers generated from an exponential distribution $p(\varepsilon) = e^{-\varepsilon/T_m}/T_m$. Here $T_m$ is an energy scale characterizing the local glassy transition for a cluster. This means that a system trapped in a cluster with energy $T_m$ will exhibit slow glassy dynamics when $T \lesssim T_m$. This energy scale $T_m$ varies from cluster to cluster and is assumed to be a random number uniformly distributed in the interval $T_m \in [0, T_f]$, where $T_f$ is the freezing temperature of the spin jam. Importantly, the energy scale for glass transition $T_f$ in a spin jam is determined by quantum fluctuations and is an order of magnitude smaller than the Curie–Weiss temperature. A similar random magnetization model is used here to describe the magnetic properties of a spin jam. Local minima within a cluster have a random magnetization $m$ uniformly distributed in the interval $[-\mathcal{M}_m, +\mathcal{M}_m]$, where the bound $\mathcal{M}_m$ itself is another random variable. Similar to the hierarchical tree counterpart, we assume a larger energy scale $T_m$ gives rise to a larger magnetization-bound $\mathcal{M}_m$.

We performed our Monte Carlo simulations following the same protocol as the experiments. The temperature decreases linearly during the cooling process, except the waiting-time period.

Numerically, we start at an initial temperature of $1.5T_f$, where $T_f \sim T_{l=1}$ is the freezing temperature, and simulate cooling by decreasing the simulation temperature in small steps ($\Delta T \sim T_f/3{,}300$). When we reach base temperature, we heat the system up (rate of $\Delta T \sim T_f/5000$). At each step, we perform 50 Monte Carlo updates. When there was waiting at an intermediate temperature, there were additional MC updates at the temperature while cooling, detailed in Fig. S2 A and B, Insets. During the reheating part of the simulations, a small magnetic field $H$ is included to generate a finite magnetization. The dc susceptibility is simply $\chi_{DC} = M/H$. The numerical results shown in Fig. 4 of the main text were obtained for different waiting temperatures after averaging over $10^5 \sim 10^7$ independent runs. Fig. S2 A and B shows the results obtained with different waiting times at $T_w = 0.6T_f$ for the spin jam and the spin glass model, respectively. Different curves in each panel correspond to varying numbers of MC steps that waited at $T_w$, which are proportional to the real waiting time $t_w$. Fig. S2 C (spin jam model) and D (spin glass model) shows $(M_{ref} - M)/M_{ref}$ as a function of (# of MC steps)/10 that best reproduces the $t_w$ (in seconds) in the experiments, as shown in Fig. 2D in the main text. In Fig. S2 C and D, $(M_{ref} - M)/M_{ref}$ is rescaled so that their maximum values are 1. As shown in Fig. S2 C and D, Insets at (# of MC steps)/10 = $10^2$, $(M_{ref} - M)/M_{ref}$ of the spin glass model is almost twice that of the spin jam model. This difference for a short waiting time is consistent with our experimental observation (Fig. 2 A–C in the main text). Furthermore, the overall dependence of $(M_{ref} - M)/M_{ref}$ as a function of $t_w$ reproduces our experimental data when scaled to the maximum value of $(M_{ref} - M)/M_{ref}$ data (Fig. 2D). We note that our MC calculations based on the multilayer random energy model do not take into account other possible sources of magnetization. As a result, different scaling factors for $(M_{re} - M)/M_{ref}$ are required to reproduce the experimental data of different systems.

The memory effect in the hierarchical tree arises from the temperature-dependent relaxation dynamics. For a given waiting temperature $T_w$, the system will fluctuate over a small window of levels depending on $T_l$ and the population $\lambda_l$ of dead-end local minima at each level. A longer waiting time $t_w$ at this temperature thus allows the system to relax to a deeper and larger (entropically) valley of the energy surface. The memory effect observed during the reheating process results from the fact that the system is trapped in this special landscape basin. Similarly, the weaker memory effect in spin jam originates from the distribution of the cluster energy scales $T_m$. With decreasing temperature $T$, thermal equilibrium cannot be reached within clusters with $T_m > T$ as the corresponding relaxation dynamics become exponentially slow. The waiting temperature sets a threshold such that clusters with $T_m > T_w$ exhibit slow relaxation dynamics, whereas a longer waiting time $t_w$ helps the system find the cluster with lower overall energy and larger entropy. Again, the fact that the system is trapped in this special cluster manifests itself as the memory effect during rewarming.

**Supplementary Figure Captions**

Fig. S1. Schematic diagram showing a Monte Carlo step in our dynamic barrier tree simulations.

Fig. S2. (A and B) The magnetization $M$ as a function of temperature $T$ during heating for the spin jam and the spin glass model, respectively. Different curves in each panel correspond to varying numbers of MC steps that waited at $T_w$, which are proportional to $t_w$. C (spin jam model) and D (spin glass model) show $(M_{ref} - M)/M_{ref}$ as a function of (# of MC steps)/10, which corresponds to tw (in seconds) in the experiments. The time scale has been chosen to best fit the experimental results. $(M_{ref} - M)/M_{ref}$ of both models are rescaled so that their maximum values are 1. As shown in C and D, Insets, at (# of MC steps)/10 = $10^2$, $(M_{ref} - M)/M_{ref}$ of the spin glass model is almost twice that of the spin jam model.

Fig. S1

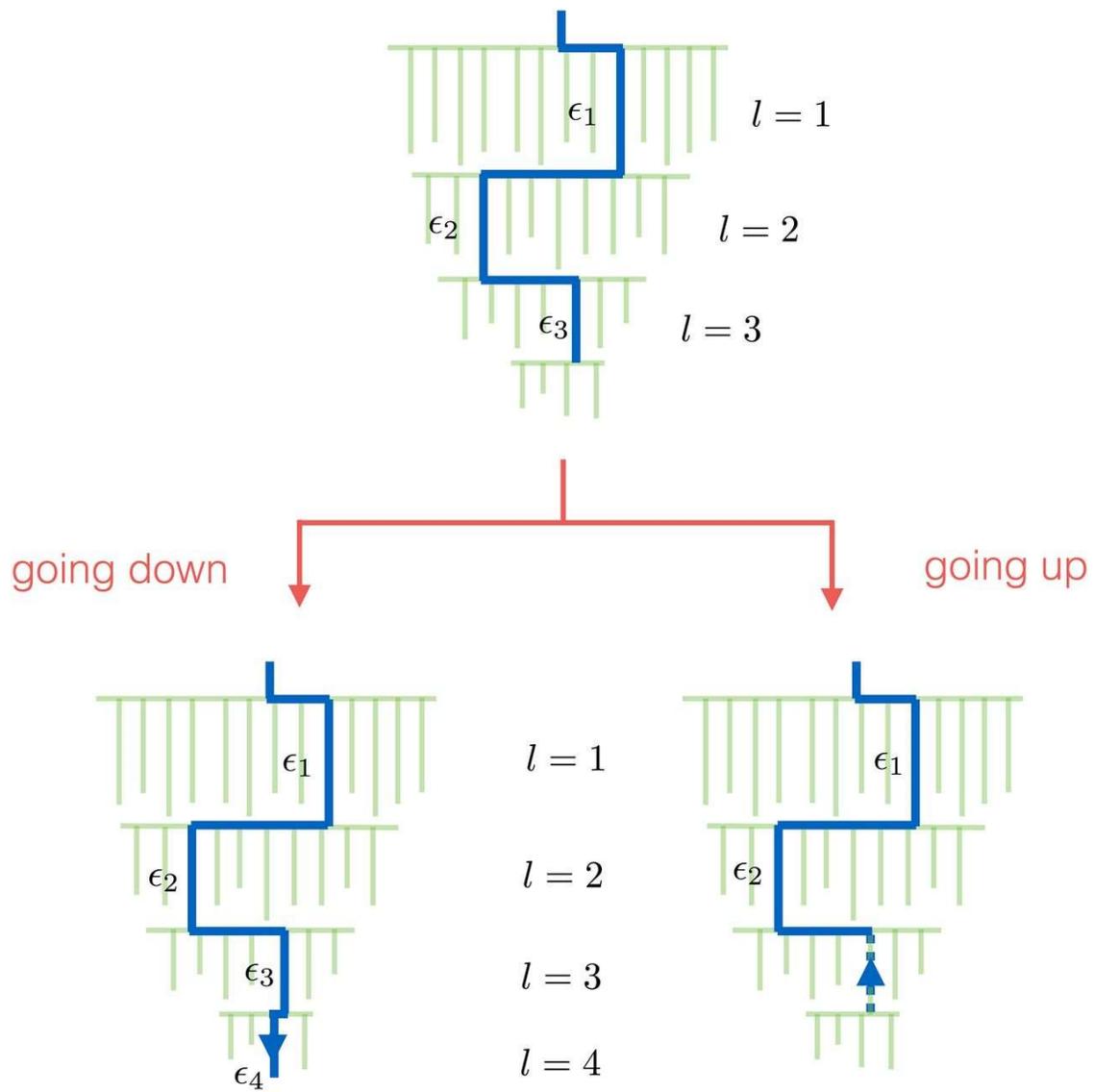

Fig. S2

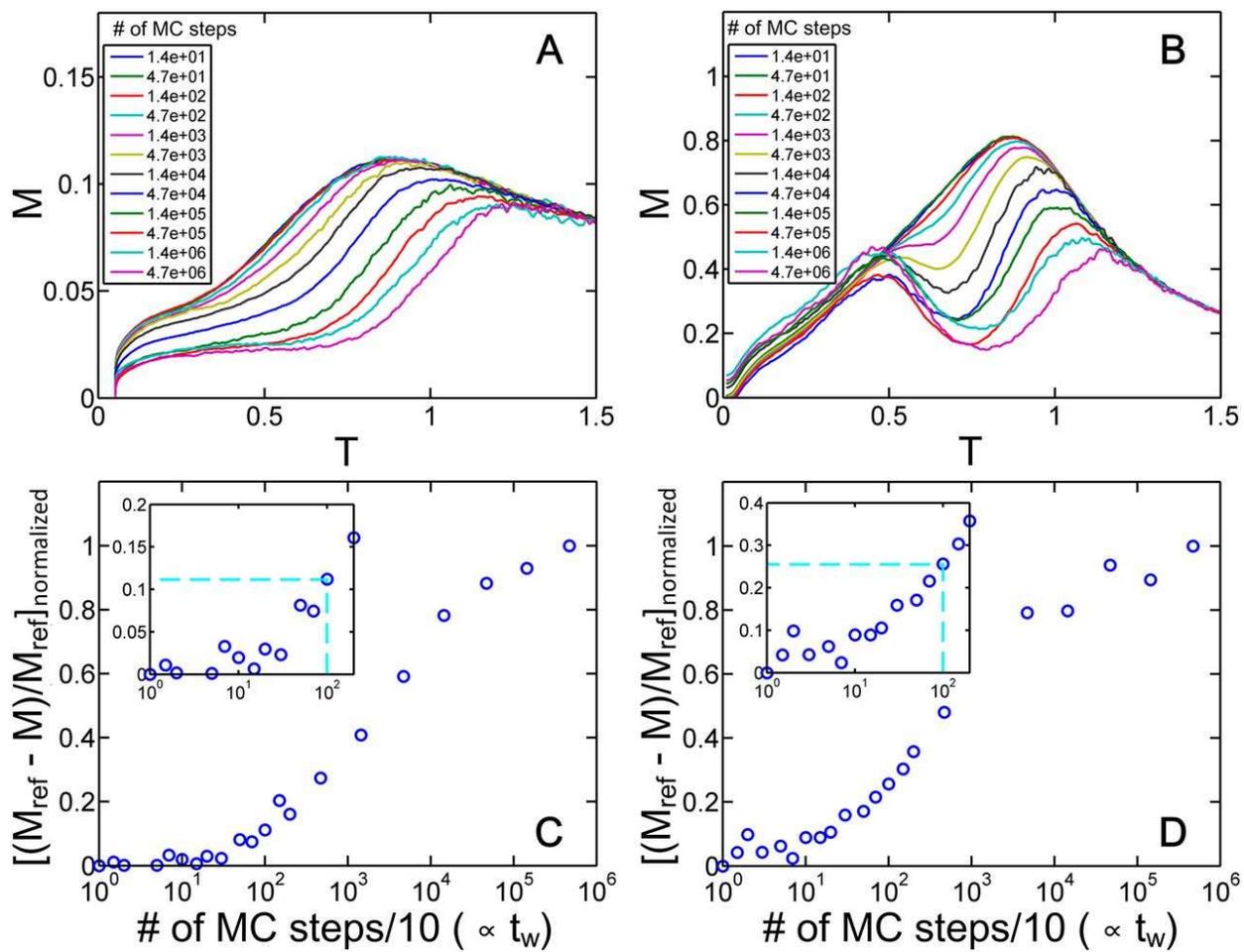